\documentclass[reprint,superscriptaddress,amsmath,amssymb,aps,twocolumn]{revtex4-1}
\usepackage{graphicx}
\usepackage{bm}
\usepackage{amsmath,amssymb,amsfonts}
\usepackage{epsfig}
\usepackage{epstopdf}
\usepackage{dcolumn}
\usepackage{grffile}
\usepackage{verbatim}
\usepackage{mathrsfs}
\usepackage{appendix}
\usepackage{xr}
\usepackage{extarrows}
\usepackage[normalem]{ulem}
\usepackage[colorlinks=true,linkcolor=blue,citecolor=blue, urlcolor=blue]{hyperref} 
    
\begin{document}

\newcommand{\fig}[2]{\includegraphics[width=#1]{#2}}
\newcommand{\la}{{\langle}}
\newcommand{\ra}{{\rangle}}
\newcommand{\dg}{{\dagger}}
\newcommand{\upa}{{\uparrow}}
\newcommand{\dna}{{\downarrow}}
\newcommand{\ab}{{\alpha\beta}}
\newcommand{\ias}{{i\alpha\sigma}}
\newcommand{\ibs}{{i\beta\sigma}}
\newcommand{\hH}{\hat{H}}
\newcommand{\hn}{\hat{n}}
\newcommand{\hc}{{\hat{\chi}}}
\newcommand{\hU}{{\hat{U}}}
\newcommand{\hV}{{\hat{V}}}
\newcommand{\br}{{\bm{r}}}
\newcommand{\bk}{{\bm{k}}}
\newcommand{\bq}{{\bm{q}}}
\newcommand{\cH}{\mathcal{H}}
\newcommand{\cG}{\mathcal{G}}
\def\gsim{~\rlap{$>$}{\lower 1.0ex\hbox{$\sim$}}}
\setlength{\unitlength}{1mm}
\newcommand{{\vhf}}{$\chi^\text{v}_f$}
\newcommand{{\vhd}}{$\chi^\text{v}_d$}
\newcommand{{\vpd}}{$\Delta^\text{v}_d$}
\newcommand{{\ved}}{$\epsilon^\text{v}_d$}
\newcommand{{\vved}}{$\varepsilon^\text{v}_d$}
\newcommand{\pprl}{Phys. Rev. Lett. \ }
\newcommand{\pprb}{Phys. Rev. {B}}
\newcommand{\LNO}{La$_3$Ni$_2$O$_7$ }

\title{Discriminating Gap Symmetries of Superconducting La$_3$Ni$_2$O$_7$}
\author{Zhan Wang}
\thanks{These authors contributed equally.}
\affiliation{Beijing National Laboratory for Condensed Matter Physics and Institute of Physics, Chinese Academy of Sciences, Beijing 100190, China}

\author{Yuxin Wang}
\thanks{These authors contributed equally.}
\affiliation{Kavli Institute for Theoretical Sciences, University of Chinese Academy of Sciences, Beijing, 100190, China}

\author{Kun Jiang}
\email{jiangkun@iphy.ac.cn}
\affiliation{Beijing National Laboratory for Condensed Matter Physics and Institute of Physics, Chinese Academy of Sciences, Beijing 100190, China}
\affiliation{School of Physical Sciences, University of Chinese Academy of Sciences, Beijing 100190, China}

\author{Jiangping Hu}
\email{jphu@iphy.ac.cn}
\affiliation{Beijing National Laboratory for Condensed Matter Physics and Institute of Physics,
Chinese Academy of Sciences, Beijing 100190, China}
\affiliation{Kavli Institute for Theoretical Sciences, University of Chinese Academy of Sciences, Beijing, 100190, China}
\affiliation{New Cornerstone Science Laboratory, 
Beijing, 100190, China}

\author{Fu-Chun Zhang}
\email{fuchun@ucas.ac.cn}
\affiliation{Kavli Institute for Theoretical Sciences, University of Chinese Academy of Sciences, Beijing, 100190, China}

\date{\today}

\begin{abstract}
   The discovery of high-T$_c$ superconductor in Ruddlesden-Popper nickelate materials represented by La$_3$Ni$_2$O$_7$ has opened new directions in the quest for unconventional superconductivity. A central unresolved issue concerns the pairing symmetry of the superconducting order. In this paper, we model the superconducting order of La$_3$Ni$_2$O$_7$ using the established Fermi surface structure together with phenomenological pairing functions belonging to $s_\pm$ and $d$-wave symmetry classes, which are the leading possibilities in the current debate. We compute several experimentally accessible observables—including tunneling density of states, point contact spectroscopy, superfluid density, and Raman spectroscopy—each of which exhibits distinct characteristics for different gap symmetries. These quantities provide a concrete and experimentally testable route for identifying the pairing symmetry of La$_3$Ni$_2$O$_7$ and for clarifying the microscopic nature of nickelate superconductivity.
\end{abstract}

\maketitle

\section{Introduction}
The discovery of superconductivity in layered nickelates has opened a new frontier in the quest for unconventional high-temperature superconductors \cite{danfeng_li,2023Wangc,2025Chene}. Among these materials, the Ruddlesden–Popper bilayer nickelate La$_3$Ni$_2$O$_7$, known as the 327 nickelate, has recently emerged as an especially intriguing platform, exhibiting superconductivity both under high pressure~\cite{2023Wangc, 2025Wangh, 2025Wangi, 2025Wang, 2023ChengCheng, chengjg_crystal, chengjg_poly, 2024Yuan, 2025Sun, 2025Huang, 2025Wend, 2025Cuia, 2025Maoa} and in thin-film form~\cite{2025Hwangc, 2025Hwang, 2025Hwangb, 2025Shen, 2025Chenb, 2025Xue, 2025Xuea, 2025Zhanga, 2025Chenc, 2025He, 2025Nie, 2025Niea, 2025Nieb, 2024Chenc, 2025Wangd, 2024Wen, 2025Wena, 2025Wenc, 2025Wange, 2025Tsukazaki}. Its rich interplay of electronic correlations, multiorbital physics, and lattice degrees of freedom parallels key features long recognized in cuprate superconductors, suggesting that the 327 nickelate family may host a new regime of strongly correlated phenomena and unconventional pairing~\cite{2025Chene}.

A central open question in this rapidly developing field is the nature of the pairing symmetry in La$_3$Ni$_2$O$_7$ superconductors~\cite{2023Zhango, 2023Zhangn, 2023Yangc, 2024Yangb, 2025Yangb, 2023Yao, 2024Yaoa, 2025Yaoa, 2023Yangb, 2024Wua, 2025Yanga, 2025Chen, 2023Si, 2024Si, 2025Si, 2023Eremin, 2024Eremin, 2025Eremin, 2025Eremina, 2025Grusdt, 2023Hub, 2025Hua, 2025Hu, 2025Han, 2023Wangd, 2024Wangg, 2023Zhangm, 2025Zhangc, 2025Jiang, 2024Zhangf, 2025Zhange, 2024Kurokia, 2025Kuroki, 2025Sakakibara, 2024Wangi, 2024Ku, 2024Wehling, 2025Wehling, 2024Lua, 2025Jiangaa, 2025Ma, 2025Wu, 2025Zhangb, 2025Chena}. Theoretically, two leading candidates have emerged: a $d$-wave ($B_{1g}$) state and a sign-changing $s_\pm$-wave ($A_{1g}$) state. Most weak-coupling approaches—driven by Fermi-surface nesting and interband scattering—favor an $s_\pm$-wave pairing~\cite{2023Wangd, 2025Eremin, 2024Wua, 2025Chen, 2023Hub, 2025Hua, 2025Hu, 2025Han, 2025Grusdt, 2025Sakakibara, 2025Li, 2025Ma, 2025Wu, 2025Chena} in which the two Fermi surfaces carry opposite signs. In contrast, strong-coupling analyses starting from a doped Mott-insulating limit tend to stabilize a cuprate-like $d$-wave gap structure~\cite{2025Zhange, 2024Xianga, 2025Li, 2025Ma, 2025Chena}. Distinguishing between these competing possibilities remains experimentally challenging, yet it is crucial for uncovering the underlying microscopic pairing mechanism.

In this work, we provide a comprehensive analysis of the experimental “smoking guns’’ capable of distinguishing between the proposed gap symmetries. Our discussion focuses on several key probes—tunneling density of states, point-contact spectroscopy, superfluid density, and Raman spectroscopy—each of which offers a distinct window into the sign structure of the superconducting order parameter. In the recently synthesized 327 nickelates, these measurements are beginning to yield important clues~\cite{2025Cuia, 2025Wend, 2025Maoa, 2025Wena, 2025He, 2025Niea, 2025Shen}, yet a unified and consistent picture of the pairing symmetry has not yet emerged. Our goal is to clarify the characteristic signatures expected for $d$-wave and $s_\pm$-wave states in these probes, thereby guiding future experiments toward a definitive resolution of the pairing structure in La$_3$Ni$_2$O$_7$.

This paper is organized as follows. We begin by introducing the tight-binding model relevant to La$_3$Ni$_2$O$_7$ and outlining the construction of various phenomenological pairing ansatzes. We then investigate the physical responses associated with each pairing state, focusing on tunneling density of states, point-contact spectroscopy, superfluid density, and Raman spectroscopy. A discussion of the implications of our findings and a summary of the main conclusions are presented in the final section.

\section{Model and Pairing Ansatz}\label{Section:model}
Since our goal is to extract the physical properties of the superconducting state, we formulate the problem within the Bogoliubov–de Gennes framework,
\begin{equation}
    \cH_{\text{BdG}}=\cH_0+\cH_\Delta^{s,d},
\end{equation}
where $\cH_0$ denotes the kinetic part derived from a tight-binding (TB) description. Because most superconducting responses are governed by the low-energy electronic structure near the Fermi level, the precise details of the TB parametrization or correlation-induced renormalizations of $\cH_0$ do not qualitatively affect our conclusions, provided the Fermi surfaces remain intact. For concreteness, we adopt TB parameters obtained from density-functional-theory (DFT) calculations \cite{2025Jiang}. The second term, $\cH_\Delta^{s,d}$, represents the superconducting pairing. As we do not address the microscopic origin of pairing in this work, we introduce phenomenological pairing ansatzes corresponding to $s_\pm$-wave and $d$-wave symmetries and specify their gap functions accordingly.

\subsection{Band Structure}
The minimal model for bilayer \LNO can be constructed using the two sets of $e_g$ orbitals residing on the top and bottom NiO$_2$ layers \cite{2023Yao}, which we label as $t$ and $b$, respectively. The tight-binding Hamiltonian takes the form
\begin{equation}
    \cH_0(\bk)=\Phi_\bk^\dagger\left(\begin{array}{cc}
    \cH_t(\bk) & \cH_\perp(\bk) \\
    \cH_\perp^\dagger(\bk) & \cH_b(\bk)
    \end{array}\right)\Phi_\bk,
    \label{eq:H0}
\end{equation}
where $\cH_b(\bk)=\cH_t(\bk)$ describes the intralayer hopping within each NiO$_2$ plane, and $\cH_\perp=\cH_\perp^\dagger$ accounts for interlayer hybridization. The basis is written as $\Phi_\bk=(c_{x,t},c_{z,t},c_{x,b},c_{z,b})^T$ with $x,z$ as orbital indices for $3d_{x^2-y^2}$ and $3d_{z^2}$, respectively. $\cH_{t,b}(\bk)$ can be written as:
\begin{equation}
    \cH_{t,b}(\bk)=\left(\begin{array}{cc}
    T_{\bk}^x & V_{\bk} \\
    V_{\bk} & T_{\bk}^z
    \end{array}\right).
\end{equation}
Here $T_{\bk}^{x/z}=t_1^{x/z}\gamma_{\bk}+t_2^{x/z}\alpha_{\bk}+\varepsilon^{x/z}$, $V_{\bk}=t_3^{xz}\beta_{\bk}$, with $\alpha_{\bk}=4\cos k_x\cos k_y$, $\beta_{\bk}=2(\cos k_x-\cos k_y)$ and $\gamma_{\bk}=2(\cos k_x+\cos k_y)$. 
$\cH_\perp$ denotes the interlayer hopping, and is written as
\begin{equation}
    \cH_\perp(\bk)=\left(\begin{array}{cc}
    t_\perp^x & V'_{\bk} \\
    V'_{\bk} & t_\perp^z
    \end{array}\right).
\end{equation}
Here $V_{\bk}'=t_4^{xz}\beta_{\bk}$. The hopping parameters are extracted from previous DFT band-structure calculations \cite{2025Jiang}, as listed in Table \ref{tab:parameter_atomic}.

\begin{figure}
    \centering
    \includegraphics[width=\linewidth]{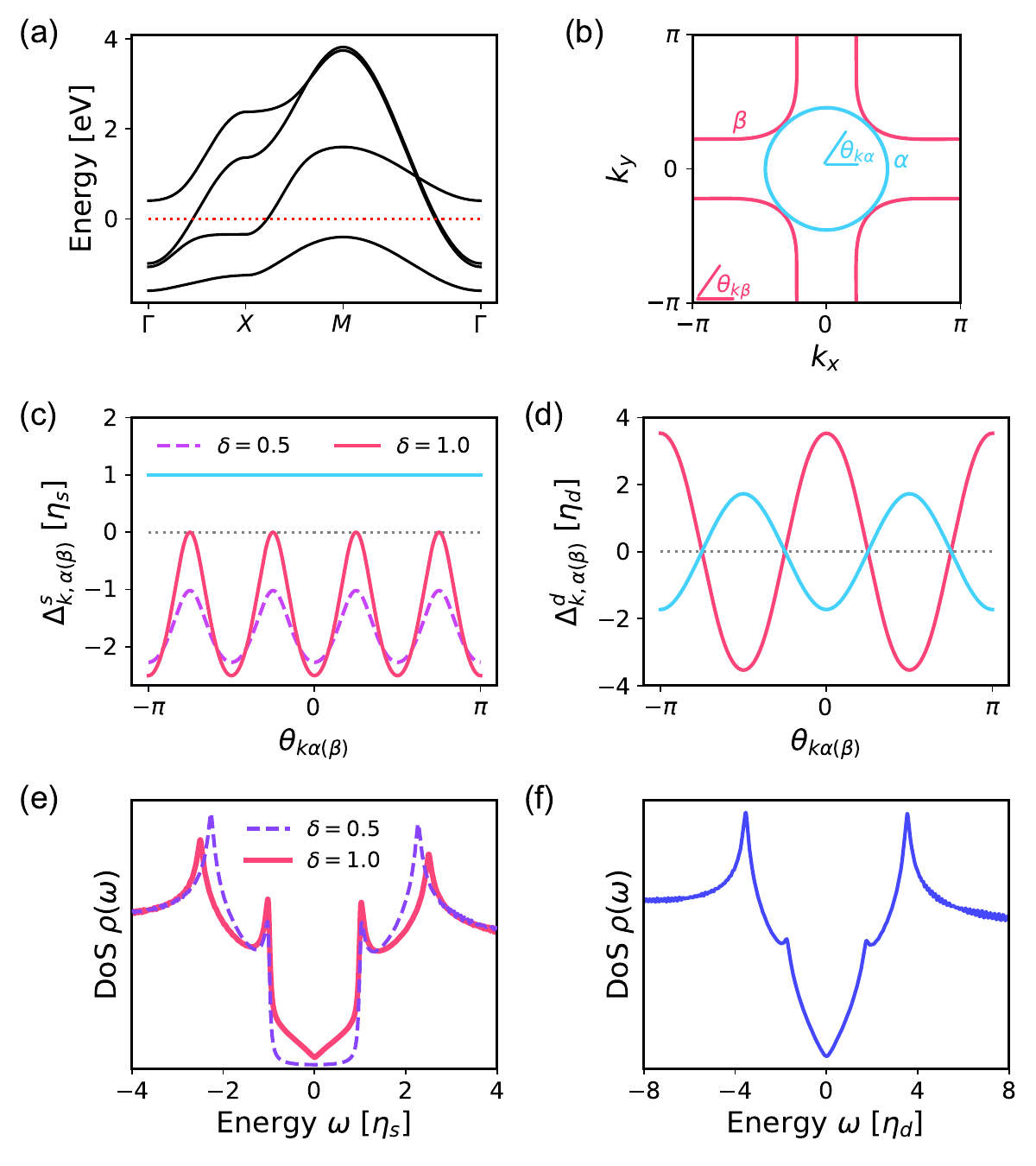}
    \caption{\textbf{Tight-binding dispersion, pairing ansatzes, and tunneling density of states.} (a) Tight-binding dispersion along the high symmetry lines in momentum space, with Fermi energy indicated by red dotted line. (b) Two Fermi surfaces obtained from the tight-binding model, where the $\alpha(\beta)$-pocket is highlighted in cyan(pink). (c, d) Two-component pairing function $\Delta_{\bk}$ projected along the two Fermi surfaces for the $s_\pm$-wave and $d$-wave pairing states, respectively, with colors corresponding to those in (b). In the case of $s_\pm$, the dashed purple (solid pink) line shows the gapped (nodal) pairing function along the $\beta$-pocket, controlled by the parameter $\delta$. (e, f) Tunneling density of states $\rho(\omega)$ calculated from the pairing function in (c, d), respectively.}
    \label{fig:basics}
\end{figure}

The energy dispersion obtained from $\cH_0(\bk)$ along high-symmetry lines in momentum space is shown in Fig.~\ref{fig:basics}(a). The system features four bands: a fully occupied $\gamma$ band, two partially filled bands labeled by $\alpha$ and $\beta$, and one unoccupied $\delta$ band. As a result, two Fermi surfaces emerge—an $\alpha$-pocket centered at the $\Gamma$ point and a $\beta$-pocket located near the Brillouin-zone (BZ) corners, as illustrated in Fig.~\ref{fig:basics}(b). There is an ongoing debate regarding whether the $\gamma$-band crosses the Fermi level \cite{2025Chene}. However, its presence does not qualitatively affect the results discussed in this work, and for simplicity, we adopt the two-Fermi-surface scenario.

\begin{table}[h]
    \centering
    \caption{The hopping parameters of the tight-binding Hamiltonian $H_0(\bk)$ in units of eV~\cite{2025Jiang}.}
    \begin{tabular}{ccccc}
    \hline
    \hline
        $t_1^x$ & $t_1^z$ & $t_2^x$ & $t_2^z$ & $t_3^{xz}$ \\
        -0.6003 & -0.149 & 0.0391 & -0.0007 & 0.2679 \\
    \hline
        $t_\perp^x$ & $t_\perp^z$ & $t_4^{xz}$ & $\varepsilon^x$ & $\varepsilon^{z}$ \\
        0.038 & -0.999 & -0.072 & 1.2193 & 0.0048 \\
    \hline
    \hline
    \end{tabular}
    \label{tab:parameter_atomic}
\end{table}

The TB Hamiltonian can also be written in terms of the mirror eigenbasis with respect to the shared apical oxygen, namely the interlayer symmetric $(+)$ and anti-symmetric $(-)$ molecular orbitals defined as:
\begin{equation}
    f_{x(z),\pm}=(c_{x(z),t}\pm c_{x(z),b})/\sqrt2.
\end{equation}
And the TB Hamiltonian can be reformulated into
\begin{equation}
    H_0(\bk)=\Psi_{\bk}^\dagger\left(\begin{array}{cc}
    \cH_+(\bk) & 0 \\
    0 & \cH_-(\bk)
    \end{array}\right)\Psi_{\bk},
\end{equation}
with $\Psi_{\bk}=(f_{\bk,x+}, f_{\bk,z+}, f_{\bk,x-}, f_{\bk,z-})^T$ and $\cH_{\pm}(\bk)=\cH_t(\bk)\pm\cH_\perp(\bk)$. The corresponding quasiparticle operators for the two partially-filled bands are denoted by $f_{\bk\alpha\sigma},\, f_{\bk\beta\sigma}$, respectively. Notably, $\alpha$-pocket belongs to the interlayer-mirror symmetric sector while $\beta$-pocket belongs to the antisymmetric sector. Because of interlayer-mirror symmetry, there is no hybridization between the two sectors. Consequently, the Wannier function associated with the two Fermi surfaces maintain distinct mirror symmetry with respect to the shared apical oxygen.

\subsection{Pairing Ansatz}
Since superconducting pairing predominantly develops on the Fermi surfaces, it is convenient to express the pairing Hamiltonian in $f_{\bk\alpha\sigma},\, f_{\bk\beta\sigma}$. As discussed above, we consider two phenomenological pairing ansatzes—$s_\pm$-wave and $d$-wave—which will be analyzed in the following.

The pairing function can be written in the form of a two-component vector $\mathbf{\Delta}_\bk^{s,d}=(\Delta_{\bk,\alpha}^{s,d},\,\Delta_{\bk,\beta}^{s,d})$, with the two components associated with the two Fermi surfaces, respectively. The general pairing Hamiltonian $H_\Delta^{s,d}$ can be written as:
\begin{equation}
    H_\Delta^{s,d}=\sum_\bk \left[\Delta_{\bk,\alpha}^{s,d} f_{\bk\alpha\uparrow}^\dagger f_{-\bk\alpha\downarrow}^\dagger + \Delta_{\bk,\beta}^{s,d} f_{\bk\beta\uparrow}^\dagger f_{-\bk\beta\downarrow}^\dagger+\text{h.c.}\right]
\end{equation}

The first pairing ansatz we consider is the $s_\pm$-wave, with the two-component pairing function in the following form:
\begin{align}
    \Delta_{\bk,\alpha}^s&=\Delta^{s0}_{\alpha},\notag\\
    \Delta_{\bk,\beta}^s&=\Delta^{s0}_{\beta}+\delta\Delta^{s1}_{\beta}\gamma_\bk^s.
\end{align}
The form factor $\gamma^s_\bk=2(\cos k_x+\cos k_y)$ corresponds to the extended $s_{\pm}$-wave component on the nearest-neighbor bonds and $\delta\in[0,1]$ controls its relative amplitude. The parameters are chosen such that the pairing function on the two components always carry opposite signs, as given in Table.~\ref{tab:parameter_pairing}. The pairing function along the Fermi surfaces are shown in Fig.~\ref{fig:basics} (c). The $\alpha$-pocket exhibits a momentum-independent $\Delta_{\bk,\alpha}^s$, plotted in cyan line, whereas the behavior of $\Delta_{\bk,\beta}^s$ is controlled by $\delta$. For $\delta=1$, the gap function $\Delta_{\bk,\beta}^s$ has four nodes (pink solid line), corresponding to a nodal $s_\pm$ state. In contrast, for $\delta<1$, exemplified by $\delta=0.5$ (purple dashed line), $\Delta_{\bk,\beta}^s$ remains fully gapped.

The other pairing ansatz considered is the $d$-wave pairing, with the pairing function specified as:
\begin{align}
    \Delta_{\bk,\alpha}^d&=\Delta^{d}_{\alpha}\gamma_\bk^d,\notag\\
    \Delta_{\bk,\beta}^d&=\Delta^{d}_{\beta}\gamma_\bk^d.
\end{align}
Here $\gamma_\bk^d=2(\cos k_x-\cos k_y)$ is the form factor for $d_{x^2-y^2}$ pairing symmetry, corresponding to pairing along the nearest neighbor bonds with alternating signs. The parameters are given in Table.~\ref{tab:parameter_pairing} and the pairing function along the two Fermi surfaces are depicted in Fig.~\ref{fig:basics} (d). In this case, both $\alpha$- and $\beta$-pockets exhibit a nodal line along the $\pm\pi/4$ direction and the sign of the pairing functions varies as a function of the angular coordinate $\theta_{\bk,\alpha(\beta)}$ within the same pocket.

\begin{table}[h]
    \centering
    \caption{Pairing parameter associated with the $s_\pm$-wave and $d$-wave pairing symmetries, with the overall amplitude set by $\eta_s$ and $\eta_d$. We use $\eta_s=0.02$ and $\eta_d=0.01$ unless stated otherwise.}
    \begin{tabular}{ccc||cc}
    \hline
    \hline
        $\Delta^{s0}_{\alpha}$ & $\Delta^{s0}_{\beta}$ & $\Delta^{s1}_{\beta}$ & $\Delta^{d}_{\alpha}$ & $\Delta^{d}_{\beta}$ \\
        $\eta_s$ & $-2.0353\eta_s$& $\eta_s$& $\eta_d$ & $\eta_d$\\
    \hline
    \hline
    \end{tabular}
    \label{tab:parameter_pairing}
\end{table}

\section{Physical Properties of the SC State}
\subsection{Tunneling Density of States}
We first calculate the tunneling density of states for different pairing ansatzes, which is defined as:
\begin{equation}
    \rho(\omega)=\sum_{\bk}\left[|u_{n\bk}|^2\delta(\omega-E_{n\bk})+|v_{n\bk}|^2\delta(\omega+E_{n\bk})\right],
\end{equation}
where $n$ denotes the BdG band index, $E_{n\bk}$ is the BdG quasiparticle energy and $u,v$ are components of the Bogolyubov quasiparticle wave functions~\cite{2016Zhu}. The tunneling density of states for a fully gapped $s_{\pm}$ pairing function and a nodal $s_\pm$ are given in Fig.~\ref{fig:basics} (e) and the $d$-wave case in Fig.~\ref{fig:basics} (f).  

All the tunneling density of states share a common feature that there are two coherent peaks, in agreement with a two-component (pocket) superconductor. On the other hand, the low-energy behaviors associated with different pairing symmetries show distinct behaviors. For a fully gapped $s_{\pm}$-wave, as depicted by dashed line in Fig.~\ref{fig:basics} (e), $\rho(\omega)$ exhibits a U-shape gap. A nodal $s_\pm$, plotted by solid line in Fig.~\ref{fig:basics} (e), maintains in-gap states with low-energy density of states following $\rho(\omega)\propto\omega^{1/2}$. $\rho(\omega)$ of the $d$-wave order maintains a linear dependence $\rho(\omega)\propto\omega$ in the low-energy regime. The tunneling density of states can be measured by the $dI/dV$ spectra by scanning tunneling microscopy (STM) for thin film \LNO samples. To identify different pairing symmetries, we propose it is crucial to distinguish the scaling behavior of $\rho(\omega)$ in the low-energy regime.

\subsection{Point-Contact Spectroscopy}

\begin{figure*}
    \centering
    \includegraphics[width=\textwidth]{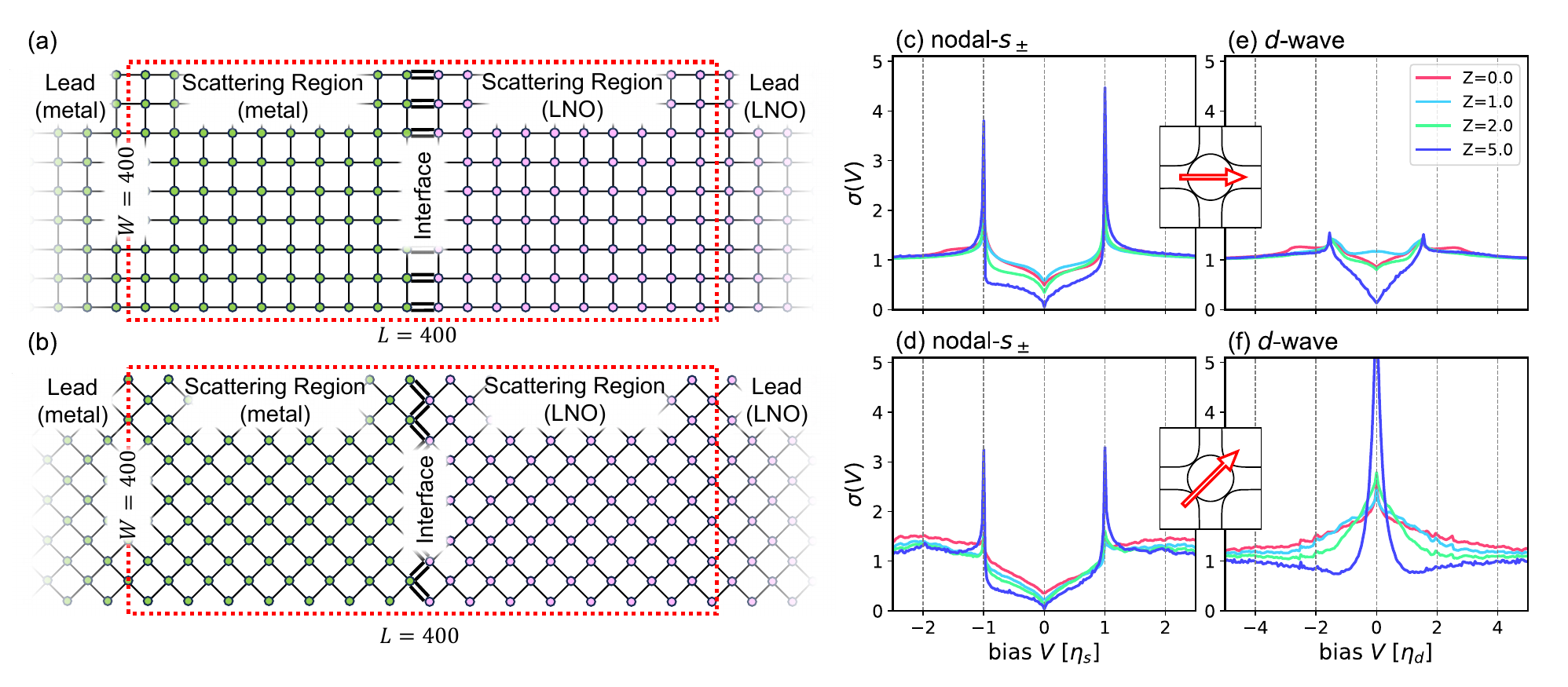}
    \caption{\textbf{Point contact setup and normalized conductance for nodal $s_\pm$-wave and $d$-wave pairing states.}  (a, b) Schematic illustration of the point-contact junction for tunneling along the (100) and (110) directions, respectively. The junction consists of two leads aligned along the $x$-direction, and a scattering region (red dashed rectangle) comprising a metallic region on the left (green sites) and \LNO on the right (pink sites). The interface contains an interfacial hopping $t_{int}$ and an onsite barrier potential $Z$ on the metallic side of the boundary, as defined in eq.~\eqref{eq:Hint}. (c-f) Normalized point-contact conductance $\sigma(V)$ calculated using KWANT. The pairing symmetry and tunneling directions are: nodal $s_\pm$ along the (100) (c) and (110) (d) , $d$-wave along the (100) (e) and (110) (f). The tunneling direction is indicated by red arrows on the insets.}
    \label{fig:PointContact}
\end{figure*}

The point contact spectroscopy (PCS) provides a direct measure to the low-energy superconducting gap structure. When a metallic tip forms a contact with a superconducting sample, the energy-dependent differential conductance spectra offers direct access to the superconducting gap structure, e.g. whether the superconducting gap is nodal or nodeless as well as the angular dependence of the gap function, making it a valuable probe to the pairing symmetry. Currently, PCS measurements are experimentally feasible to pressurized \LNO but a clear conclusion for the pairing symmetry is still under debate~\cite{2025Wend, 2025Cuia, 2025Maoa}.

In our study, PCS is calculated using the KWANT package \cite{kwant}. Specifically, the transmission rate $T(V)$, as a function of the bias voltage $V$, is calculated and it is related to the differential conductance $G(V)$ via the Landauer-Buttiker equation:
\begin{equation}
    G(V)=\frac{e^2}{\hbar}T(V).
\end{equation}
The setup in our calculation contains two leads and a scattering region. Electrons can tunnel through the junction along the (100) and the (110) directions, as schematically plotted in Fig.~\ref{fig:PointContact} (a) and (b), respectively. 

The left lead as well as the left part of the scattering region, describing a metallic wire, is modeled by the Hamiltonian $H_{metal}$:
\begin{equation}
    H_{metal}=-\sum_{\langle ij\rangle_n\sigma}\left[t_{n}d_{i\sigma}^\dagger d_{j\sigma} +\text{h.c.}\right]-\mu_m\sum_{i\sigma}d_{i\sigma}^\dagger d_{i\sigma},
\end{equation}
where $\langle ij\rangle_n$ denotes $n$-th nearest neighbor bonds on a square lattice. Specifically, we consider $t_1=1.0$ for nearest neighbors and $t_2=0.35$ for 2nd nearest neighbors. The chemical potential on the metallic side is set at $\mu_\text{m}=0$. For the right part, we consider a two-band model for the \LNO superconductor with electron operators denoted as $f_{i,\alpha(\beta),\sigma}$, corresponding to the quasiparticles in $\alpha$ and $\beta$ bands. The Fermi surface profile and the pairing functions are the same as introduced in the last section.

The metal-superconductor interface consists of a interfacial hopping $t_{int}$ and a barrier potential $Z$ on the metal side of the interface. The Hamiltonian can be denoted as:
\begin{align}
    H_{int}=&-\sum_{\langle i_mj_s\rangle,\sigma} t_{int} (d_{i_m\sigma}^\dagger f_{j_s\alpha\sigma} + d_{i_m\sigma}^\dagger f_{j_s\beta\sigma}+\text{h.c.})\notag\\
    &+\sum_{i_m,\sigma} Z d_{i_m\sigma}^\dagger d_{i_m\sigma},
    \label{eq:Hint}
\end{align}
where $i_{m}(j_s)$ denotes the site index on the metal (superconductor) side and $\langle i_m j_s\rangle$ denotes the interfacial nearest-neighbor bond.

In calculation, we consider the scattering region with size $W=L=400$, with open boundary condition along the $y$-direction. The two ends along the $x$-direction are connected to the metallic and superconducting leads, respectively. The interface is set at the middle of the scattering region, as shown in Fig.~\ref{fig:PointContact} (a, b). The tunneling hopping amplitude in eq.~\eqref{eq:Hint} is set at  $t_{int}=0.5$eV and the barrier energy varies from $Z=0$eV to $Z=5$eV. We calculate the normalized conductance, defined as $\sigma(V)=G_S(V)/G_N(V)$ where $G_{S(N)}$ is the differential conductance of the (non)-superconducting differential conductance. In calculation of $G_N$, the pairing functions on the \LNO side are set to zero.

The results for normalized conductance $\sigma(V)$ is presented in Fig.~\ref{fig:PointContact} (c-f). For both pairing symmetries, with increasing barrier height $Z$, the normalized conductances show similar behaviors with the features becoming more pronounced for larger $Z$ because Andreev reflections are further suppressed. The directional dependence of the normalized conductance shows distinct features between the nodal $s_\pm$ and the $d$-wave pairing ansatzes. For nodal $s_\pm$ with electrons tunneling along the (100) direction, as plotted in Fig.~\ref{fig:PointContact} (c), $\sigma(V)$ maintains coherent peaks at bias voltage $V=\pm\eta_s$, which originates from the superconducting gap along the $\alpha$-pocket. The coherent peaks from the $\beta$-pocket, however, is absent due to the particular choice of Fermi surface on the metallic side. Moreover, in the low-energy regime, we obtain $\sigma(V)\propto|V|^{1/2}$ behavior, similar to the tunneling density of states shown in Fig.~\ref{fig:basics} (e). For electrons tunneling along the (110) direction, shown in Fig.~\ref{fig:PointContact} (d), similar behaviors are found as the tunneling along the (100) direction, including coherent peaks at $V=\pm\eta_s$ and $\sigma(V)\propto|V|^{1/2}$ gap behaviors in low-bias region. 

In the $d$-wave case, the normalized conductance along (100), as depicted in Fig.~\ref{fig:PointContact} (e), shows a $V$-shape gap with small coherent peaks from the superconducting gap of the $\alpha$-pocket. The coherent peaks from the $\beta$-pocket is also absent for the same reason as in the nodal $s_\pm$ case. Along the (110) direction, on the other hand, $\sigma(V)$ is found to have a strong zero-bias peak at $V=0$, as shown in Fig.~\ref{fig:PointContact} (f). 

The strong directional dependence of the $d$-wave conductance originates from the destructive interference of the alternating signs in the pairing function~\cite{1995Kashiwaya, 2022Xiang}. Specifically, the sign of the pairing function is angular-dependent within the same pocket, leading to different behaviors for tunneling along the (100) and (110) directions. In contrast, the sign change of the nodal $s_\pm$ is between the two Fermi pockets, while the sign of the pairing functions are rotation-invariant within the same pocket. This can explain why the destructive interference for $s_\pm$ considered here is direction-independent, resulting in similar results for (100) and (110) tunnelings. To this end, we propose the directional contrast in the PCS can provide a diagnostic evidence to resolve different pairing symmetries.

\subsection{Superfluid density}
Superfluid density $n_S(T)$ is a fundamental quantity that reflects the system’s ability to maintain long-range phase coherence, thereby linking microscopic pairing mechanisms to macroscopic electromagnetic responses such as the London penetration depth $\lambda(T)$ and the Meissner effect. 

\begin{figure}
    \centering
    \includegraphics[width=\linewidth]{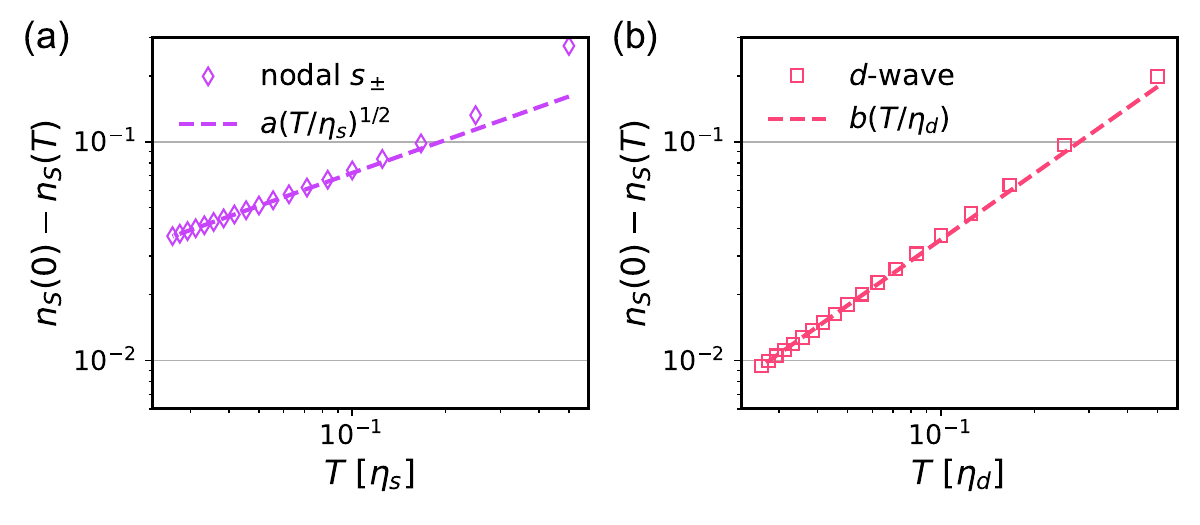}
    \caption{\textbf{Superfluid density calculated for nodal $s_\pm$-wave and $d$-wave pairing ansatzes.} (a) Decay in superfluid density $n_S(0)-n_S(T)$ for nodal $s_\pm$-wave, showing a $a(T/\eta_s)^{1/2}$ behavior, as fitted by the dashed line with $a\approx1.14$. (b) $n_S(0)-n_S(T)$ obtained with $d$-wave ansatz, showing a linear in $bT/\eta_d$ decay, as fitting by the red dashed line with $b\approx8.93$. Parameter: $\eta_s=\eta_d=\eta=0.04$eV.}
    \label{fig:SCdensity}
\end{figure}

Experimentally, superfluid density $n_S(T)$ can be measured by the magnetic penetration depth $\lambda(T)$ following the relation $n_S(T)\propto\lambda^{-2}(T)$, and serves as a key feature of the superconducting gap. Revealing the distinct temperature dependence of the superfluid density therefore provides important evidence for the pairing symmetry in unconventional superconductors, e.g. in cuprates~\cite{1993Zhang, 2022Xiang} and infinite-layer nickelates~\cite{2025Hwangi, 2023Ariando}.

Theoretically, the superfluid density $n_S(T)$ can be calculated by taking the real part of the current response function $K_{\mu\nu}(\bq,\omega)$ in the limit $\omega=0$ and $\bq\rightarrow0$, i.e.
\begin{equation}
    n_{S,\mu}(T)=\Re[K_{\mu\mu}(\bq\rightarrow0,\omega=0)],
\end{equation}
where $\mu,\nu$ denotes the spatial direction and $K_{\mu\nu}(\bq,\omega)$ characterizes the current response $J_\mu(\bq,\omega)$ to an external electromagnetic field $A_\nu(\bq,\omega)$ according to $J_\mu(\bq,\omega)=-\sum_\nu K_{\mu\nu}(\bq,\omega)A_\nu(\bq,\omega)$. For \LNO, we consider the in-plane superfluid density with $C_4$ symmetry, i.e. $n_{S,x}(T)=n_{S,y}(T)=n_S(T)$. The calculation details are written in Appendix.~\ref{Appendix:SD}. To identify different pairing symmetries, we explicitly calculate the superfluid density by considering the $s_\pm$ and $d$-wave pairing symmetries and identify their low-temperature behaviors at $T\ll\Delta$.

The superfluid density decay, $n_S(0)-n_S(T)$, calculated with respect to the extracted limit at $T=0$, is shown in Fig.~\ref{fig:SCdensity}. In the case of nodal $s_\pm$-wave, the superfluid density decay is found to follow the square root of the temperature, i.e.
\begin{equation}
    n_S(0)-n_S(T)\propto a(T/\eta_s)^{1/2},
\end{equation}
where $a$ is some coefficient. The relation is revealed by fitting the data points using the dashed line in Fig.~\ref{fig:SCdensity} (a). On the other hand, $n_S(T)$ of a fully gapped $s_\pm$-wave with $\delta=0.5$, corresponding to the SC gap structure plotted as dashed line in Fig.~\ref{fig:basics} (c), is found to decay exponentially as a function of temperature. Superfluid density of a $d$-wave SC order, as shown in Fig.~\ref{fig:SCdensity} (b), is found to decay linearly with temperature, 
\begin{equation}
    n_S(0)-n_S(T)\propto b(T/\eta_d),
\end{equation}
where $b$ is some constant coefficient. 

The low-temperature behaviors for different pairing symmetries follows closely from the low-energy scaling relation of the density of states $\rho(\omega)$. In particular, the decay of superfluid density can be estimated according to:
\begin{equation}
    n_S(0)-n_S(T)\propto\int d\omega \rho(\omega)\partial n_F(\omega)/\partial \omega,
\end{equation}
where $n_F(\omega)$ is the Fermi distribution function. For nodal $s_\pm$, we have $\rho(\omega)\propto|\omega|^{1/2}$ in low energy region. For $T\ll\eta_s$ the superconducting gap, the integral is restricted to $\omega<T$ and is estimated to be proportional to $T^{1/2}$. In comparison, the low-energy density of states for $d$-wave superconductor scales as $\rho(\omega)\propto|\omega|$ and the superfluid density follows a linear in $T$ decay~\cite{1993Zhang, 2022Xiang}.
We therefore propose the measurement of superfluid density to be another signature to identify different pairing symmetries.

\subsection{Raman spectroscopy}
Raman spectroscopy provides another powerful probe of superconducting pairing symmetry. As an optical technique based on inelastic photon scattering, it offers a symmetry-resolved view of low-energy electronic dynamics. The Raman intensity is proportional to the differential scattering cross section \cite{2022Xiang},
\begin{align}
    \frac{\partial^{2}\sigma}{\partial\Omega\partial\omega}\propto-\frac{1}{1-e^{-\beta\omega}}\mathrm{Im}R(\bq, \omega),
\end{align}
where $R(\bq, \omega)$ is the Raman response function. In the time domain, this response is expressed as the correlation function of the Raman density operator $\rho_{\gamma}$,
\begin{align}
    R(\bq, \omega)=-i\theta(t)\langle[\rho_{\gamma}(\bq, t), \rho_{\gamma}(-\bq, 0)]\rangle.
\end{align}
For optical experiments, the momentum transfer is negligible, so we take $\mathbf{q}\to 0$. And the Raman density operator becomes
\begin{align}
    \rho_{\gamma}(0, t)=\sum_{\bk,m,n,\sigma}\gamma_{\bk}^{mn}c_{\bk,m,\sigma}^{\dagger}(t)c_{\bk,n,\sigma}(t),
\end{align}
where $m, n$ label orbitals and $\sigma$ is the spin index. The Raman vertex $\gamma_{\mathbf{k}}$ is defined by \cite{devereaux2007inelastic}
\begin{align}
    \gamma_{\bk}=\sum_{\mu,\nu}e_{\mu}^{I}\frac{\partial^2H_0}{\partial k_{\mu}\partial k_{\nu}}e_{\nu}^{S},\label{gamma}
\end{align}
with $e^{I}$ and $e^{S}$ the polarization vectors of the incident and scattered light, $H_0$ defined in eq.~\eqref{eq:H0} and $\mu,\nu = x,y,z$ labeling Cartesian coordinates. Additional technical details regarding the Raman response are presented in the Appendix B.

\begin{figure}
    \centering
    \includegraphics[width=\linewidth]{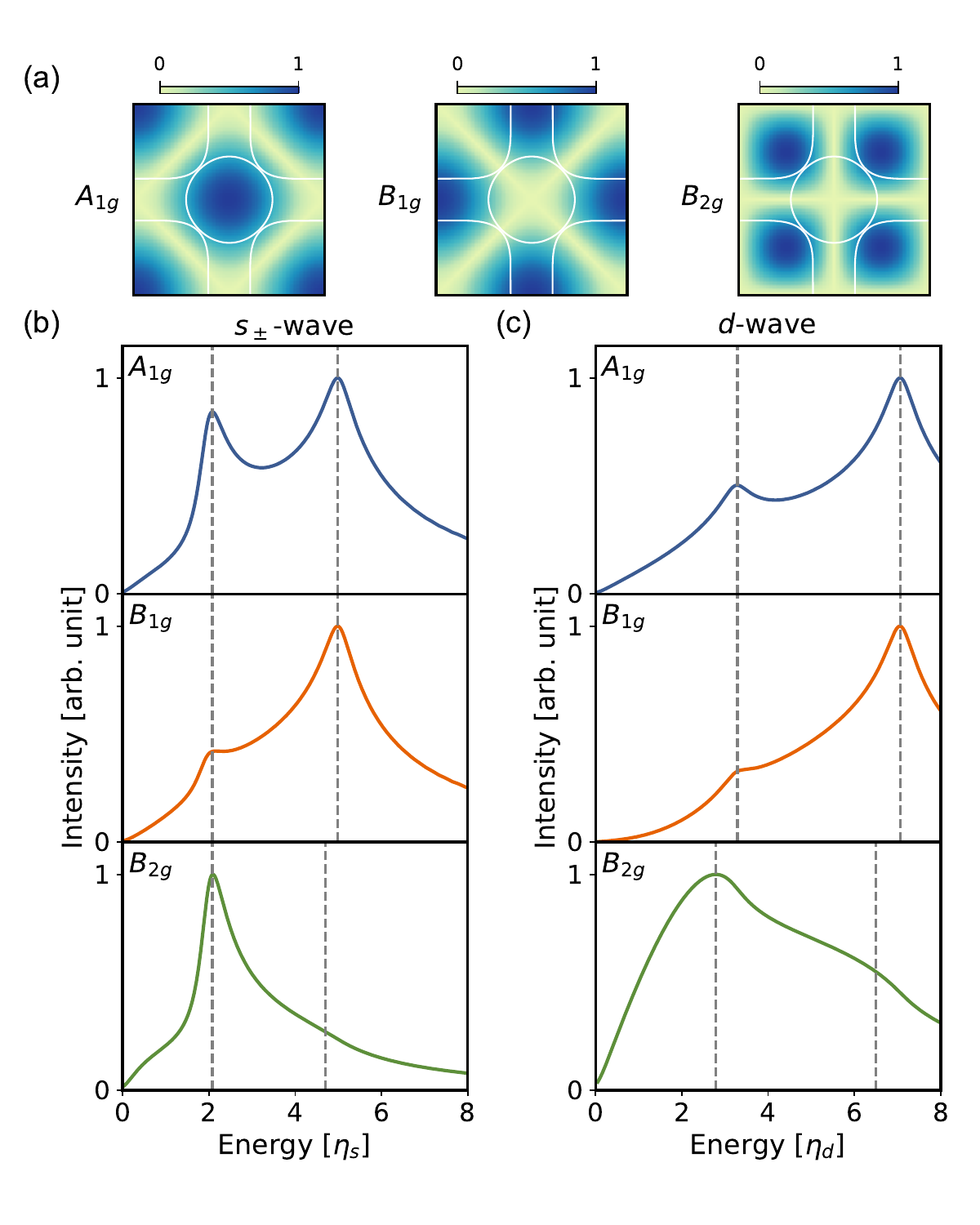}
    \caption{\textbf{Raman spectra of nodal $s_\pm$-wave and $d$-wave pairing in  $A_{1g}$, $B_{1g}$, and $B_{2g}$ channels.} (a) The diagram illustrating the detection of SC gaps at different positions in the BZ in $A_{1g}$, $B_{1g}$, and $B_{2g}$ channels, respectively. The blue areas approximately represent the regions that can be detected. The white line is the Fermi surface. (b) The Raman spectra obtained from three different symmetry types of light fields when the symmetry of the SC gap is nodal $s_{\pm}$-wave. (c) The Raman spectra obtained from three different symmetry types of light fields when the symmetry of the SC gap is $d$-wave. The gray dashed line represents the position of the peak. When the SC gap in the detection region is small, the corresponding Raman spectrum peaks shift to the lower energy.}
    \label{fig:Raman}
\end{figure}

By selecting different combinations of incident and scattered light polarizations, Raman spectroscopy can be tuned to probe distinct regions of the Fermi surface. As indicated by Eq.~\ref{gamma}, the polarization vectors directly determine the structure of the Raman vertex $\gamma_{\mathbf{k}}$, which acts as a form factor weighting different $\mathbf{k}$-space regions in the Raman response. In our analysis, we consider three standard symmetry channels: $A_{1g}$, $B_{1g}$, and $B_{2g}$. 
 For the $B_{1g}$ channel, choosing $e^{I}=(\frac{\sqrt{2}}{2}, \frac{\sqrt{2}}{2})$ and $e^{S}=(\frac{\sqrt{2}}{2}, -\frac{\sqrt{2}}{2})$ produces
$\gamma_{\bk}^{B_{1g}}=\frac{1}{2}(\frac{\partial^{2}H_0}{\partial k_{x}^{2}}-\frac{\partial^{2}H_0}{\partial k_{y}^{2}})$. For the $B_{2g}$ channel, choosing $e^{I}=(1,0)$ and $e^{S}=(0,1)$ yields
$\gamma_{\bk}^{B_{2g}}=\frac{\partial^{2}H_0}{\partial k_{x} \partial k_{y}}$ \cite{2022Xiang,devereaux2007inelastic}. The $A_{1g}$ component,
$\gamma_{\bk}^{A_{1g}}=\frac{1}{2}(\frac{\partial^{2}H_0}{\partial k_{x}^{2}}+\frac{\partial^{2}H_0}{\partial k_{y}^{2}})$
cannot be isolated from a single polarization geometry; instead, it is obtained by combining measurements from a symmetric configuration and subtracting the $B_{2g}$ contribution \cite{2022Xiang}.
Fig.~\ref{fig:Raman}(a) illustrates their corresponding $\gamma_{\mathbf{k}}$ patterns, overlaid with a representative Fermi surface to highlight which momentum regions each channel emphasizes. The fully symmetric $A_{1g}$ channel samples the Brillouin zone uniformly; $B_{1g}$ focuses on the anti-nodal $(\pi,0)$ directions; and $B_{2g}$ highlights the nodal $(\pi,\pi)$ diagonals. Because the $B_{1g}$ and $B_{2g}$ vertices emphasize anti-nodal and nodal regions respectively, they are particularly sensitive to the contrast between $s_{\pm}$-wave and $d$-wave gaps. In a superconductor where these two directions host markedly different gap magnitudes or nodes, one naturally expects the corresponding Raman spectra in the $B_{1g}$ and $B_{2g}$ channels to exhibit clear, symmetry-selective distinctions.

In our calculation, we set $\beta$ = 1000 and employed a $\bk$-mesh of 1500 × 1500. As can be seen from Figs. \ref{fig:Raman}(b, c), the Raman spectra exhibit entirely distinctive characteristics for $s_{\pm}$-wave and $d$-wave pairings. The $A_{1g}$ spectrum, representing a global average, provides a benchmark for identifying the number and average positions of peaks. Owing to the presence of only two Fermi surfaces in our employed TB model, only two peaks are consequently expected for both $s_{\pm}$-wave and $d$-wave pairing symmetries. The gap magnitudes displayed in Figs. \ref{fig:basics}(c, d) allow for the assignment of the low-frequency peak to the $\alpha$ Fermi surface and the high-frequency peak to the $\beta$ Fermi surface. The other two symmetries exhibit strong directional selectivity. If the probed direction aligns with a gap minimum (or a node), then the corresponding peak will exhibit a leftward shift \cite{2022Xiang}. For nodal $s_\pm$-wave case, the gap magnitude on the $\alpha$ Fermi surface is isotropic. Consequently, the low-frequency peaks for all three symmetries are located at the same position. On the other hand, the gap on the $\beta$ Fermi surface has a node in the nodal direction and reaches its maximum in the anti-nodal direction. As a result, the high-frequency peak for $B_{1g}$ coincides with that for $A_{1g}$, while the $B_{2g}$ high-frequency peak (indicated by the gray dashed line) is shifted to a lower frequency in Fig. \ref{fig:Raman}(b). The situation for $d$-wave pairing is analogous. Both the $\alpha$ and $\beta$ Fermi surfaces possess nodes in the nodal direction and exhibit maximum gaps in the anti-nodal direction. Consequently, the positions of the two peaks in the $B_{1g}$ spectrum should coincide with those in the $A_{1g}$ spectrum, while both peaks in the $B_{2g}$ spectrum (indicated by gray dashed line) are expected to shift to the left in Fig. \ref{fig:Raman}(c). 

\section{Discussion}

Although our goal is to identify clear physical distinctions between the proposed $s_{\pm}$-wave and $d$-wave pairing states in \LNO, reaching a definitive conclusion remains challenging at the present stage. Two key factors contribute to this difficulty: the limited quality of currently available samples and the narrow range of experimental probes that can be reliably applied. The synthesis of high-quality nickelate samples—whether thin films or single crystals—remains a major bottleneck for the field.

For the high-pressure superconducting state, transport measurements are the most accessible and have provided the first indications of unconventional behavior. Point-contact spectroscopy offers a more direct probe of the gap structure. Optical techniques, including optical conductivity and Raman scattering, are also feasible under pressure and merit further systematic exploration.

In thin-film systems, a broader suite of experimental tools becomes available. High-resolution angle-resolved photoemission spectroscopy (ARPES) and STM can directly probe gap anisotropy, coherence peaks, and potential nodal behavior in sufficiently clean samples. Additionally, measurements of the superfluid density—already carried out in infinite-layer nickelates~\cite{2025Hwangi, 2023Ariando}—provide complementary constraints that can help distinguish between competing pairing symmetries. Taken together, these approaches highlight both the opportunities and the current limitations in conclusively determining the superconducting order parameter of \LNO.

The response of a superconducting state to nonmagnetic impurities also provides clues to its underlying pairing symmetry. As discussed in a recent work~\cite{2025Eremina}, the $d$-wave order is suppressed with nonmagnetic impurities, while the impact on the $s$-wave is more subtle to the electronic structures. A systematic understanding of the disorder effects on the pairing symmetry and a concrete relation to the chemical properties in realistic materials remain to be established with more theoretical and experimental inputs.

Experimentally, a variety of spectroscopic probes have begun to reveal the superconducting gap structure in \LNO. ARPES measurements report a nearly momentum-homogeneous superconducting gap~\cite{2025He, 2025Niea}. STM studies reveal the presence of two distinct superconducting gap scales, suggesting multiband superconductivity \cite{2025Wena}. PCS measurements have also been carried out, with the existence of a zero-bias peak still under debate~\cite{2025Cuia, 2025Wend, 2025Maoa}.
Together, these experimental results highlight both the progress and the current ambiguity in resolving the pairing symmetry of \LNO.

Before concluding, we note that the microscopic pairing mechanism does not affect the low-energy signatures discussed in this work. Whether the superconductivity in \LNO originates from weak-coupling interactions, strong-coupling physics, or more complex multiorbital processes, the resulting low-energy excitations are governed primarily by the gap symmetry and the topology of the Fermi surfaces. This insensitivity to microscopic details is well established in the study of cuprate $d$-wave superconductors. A comprehensive summary of this principle can be found in~\cite{2022Xiang}, which shows that the gap symmetry dictates the observable low-energy properties.

\section{Summary}
The discovery of \LNO marks a crucial advancement in the nickel age of the high-T$_c$ superconductivity. The unique bilayer structure and multiple electronic orbitals further enriches the understanding of unconventional superconductivity. Identifying the pairing symmetry is a central step in characterizing the superconducting order. In this work, we analyze several key experimental observables that serve as smoking-gun signatures capable of distinguishing between candidate pairing states, ranging from the $s_\pm$-type $A_{1g}$ representation to the $d$-wave $B_{1g}$ representation. Using phenomenological gap functions to model each superconducting state, we compute physical properties including the tunneling density of states, point contact spectroscopy, superfluid density measured by London penetration depth, and Raman spectroscopy. These quantities exhibit distinct and characteristic behaviors for each pairing symmetry, providing experimentally accessible criteria to discriminate between the proposed superconducting states.
We hope that our results will further motivate and guide experimental efforts to determine the pairing symmetry in \LNO.

\section{Acknowledgement}
 We acknowledge the support by the National Natural Science Foundation of China (Grant NSFC-12494594, NSFC-12574150, NSFC-12174428), the Ministry of Science and Technology (Grant No. 2022YFA1403900), the Chinese Academy of Sciences Project for Young Scientists in Basic Research (2022YSBR-048), the Innovation program for Quantum Science and Technology (Grant No. 2021ZD0302500), Chinese Academy of Sciences under contract No. JZHKYPT-2021-08, and the New Cornerstone Investigator
Program.

\bibliography{ni327-SC-properties}

\clearpage

\appendix
\onecolumngrid

\section{superfluid density}\label{Appendix:SD}
We start from a multi-orbital BdG Hamiltonian written as:
\begin{equation}
    H_{BdG}=H_0+H_\Delta,
\end{equation}
where $H_{0,\Delta}$ denote the kinetic and pairing part, respectively. For a multiorbital Hamiltonian, we use $\alpha,\beta$ to denote the orbital indices and the kinetic Hamiltonian takes the general form:
\begin{equation}
    \cH_0=-\sum_{\langle ij\rangle,\sigma}\sum_{\alpha\beta} (t_{ij}^{\alpha\beta} c_{i\alpha\sigma}^\dagger c_{j\beta\sigma} + \text{h.c.}) =\sum_{\bk,\sigma}\sum_{\alpha\beta}\varepsilon_\bk^{\alpha\beta}c_{\bk\alpha\sigma}^\dagger c_{\bk\beta\sigma}.
\end{equation}
Here $\varepsilon_\bk^{\alpha\beta}=-\sum_{\langle ij\rangle}2t_{ij}^{\alpha\beta}\cos \bk\cdot \br_{ij}$ is the corresponding matrix element. As for $\cH_\Delta$, we consider the various rages including onsite pairing and pairing between 1st nearest neighbors.

The current response to an external electromagnetic field is described by the Kubo formula:
\begin{equation}
    J_\mu(\bq,\omega)=-\sum_\nu K_{\mu\nu}(\bq,\omega)A_\nu(\bq,\omega).
\end{equation}
Here $K_{\mu\nu}(\bq,\omega)$ is the conductivity tensor. The superfluid density $n_{S,\mu}$ is obtained by taking the limit $\bq\rightarrow0$ and $\omega=0$ of the diagonal element of the conductivity tensor, i.e. 
\begin{equation}
    n_{S,\mu}=K_{\mu\mu}(\bq\rightarrow0,\omega=0).
\end{equation}

The response function $K_{\mu\nu}$ is determined by the electron effective mass and the current-current correlation function:
\begin{equation}
    K_{\mu\nu}(\bq,\omega)=\frac{e^2}{V\hbar^2} \sum_{\bk} \langle\frac{\partial H_0}{\partial k_\mu\partial k_\nu}\rangle + \Pi_{\mu\nu}(\bq,\omega).
\end{equation}

The first term represents the diamagnetic contribution and comes from the effective mass of the quasi-particle, which can be evaluated as:
\begin{equation}
    \frac{e^2}{V\hbar^2} \sum_{\bk} \langle\frac{\partial H_0}{\partial k_\mu\partial k_\nu}\rangle=\frac{e^2}{V\hbar^2} \sum_{\bk\sigma}\sum_{\alpha\beta} \frac{\partial \varepsilon_\bk^{\alpha\beta}}{\partial k_\mu\partial k_\nu}\langle c_{\bk\alpha\sigma}^\dagger c_{\bk\beta\sigma}\rangle.
\end{equation}
Here the expectation value can be calculated under finite temperature in the superconducting state.

The second term is the paramagnetic contribution. $\Pi_{\mu\nu}(\bq,\omega)$ is the current-current correlation function defined as:
\begin{equation}
    \Pi_{\mu\nu}(\bq,\omega)=-\frac{i}{V\hbar^2}\int_0^\infty dt e^{i\omega t}\langle[J_\mu(\bq,t),J_\nu(-\bq,0)]\rangle.
\end{equation}
The current operator is defined as:
\begin{equation}
    J_\mu(\bq)=e\sum_{\bk\sigma} \sum_{\alpha\beta} \frac{\partial \varepsilon^{\alpha\beta}_{\bk+\bq/2}}{\partial k_\mu} c_{\bk+\bq,\alpha,\sigma}^\dagger c_{\bk\beta\sigma}.
\end{equation}
Note we use $e<0$ to denote the electron charge.

In the superconducting state, we adopt the Nambu basis $\Phi_\bk=(c_{\bk\alpha\uparrow}, c_{-\bk\alpha\downarrow}^\dagger)$ with $\alpha$ denoting all orbitals. The current operator is written as:
\begin{equation}
    J_\mu(\bq)=e\sum_\bk\sum_{\alpha\beta} \frac{\partial \varepsilon^{\alpha\beta}_{\bk+\bq/2}}{\partial k_\mu} (c_{\bk+\bq,\alpha,\uparrow}^\dagger c_{\bk\beta\uparrow} + c_{-\bk-\bq,\beta,\uparrow} c_{-\bk\alpha\uparrow}^\dagger),
\end{equation}
where we make use of the properties that $\varepsilon_{\bk}^{\alpha\beta}=\varepsilon_{\bk}^{\alpha\beta}$ is an even function of $\bk$ and is orbital-symmetric $\varepsilon_{\bk}^{\alpha\beta}=\varepsilon_{\bk}^{\beta\alpha}$ in the current study. By defining $\Xi_{\bk,\mu}=\frac{\partial \varepsilon^{\alpha\beta}_{\bk}}{\partial k_\mu}\otimes\tau_0$ with $\tau_0$ being the identity matrix in the particle-hole space, we get the form of $J_\mu(\bq)$:
\begin{equation}
    J_\mu(\bq)=e\sum_\bk\Phi_{\bk+\bq}^\dagger\Xi_{\bk+\bq/2}\Phi_\bk.
\end{equation}
We evaluate the current-current correlation function in the imaginary time $\tau$ and perform analytical continuation to obtain its real frequency dependence. Using the Gorkov Green's function defined by the Nambu spinors:
\begin{equation}
    \cG(\bk,i\omega_m)=-\int_0^\beta e^{i\omega_m\tau} \langle \Phi_\bk(\tau)\Phi_\bk^\dagger(0)\rangle d\tau=\left(i\omega_m-\cH_{BdG,\bk}\right)^{-1},
\end{equation}
where $\omega_m$ is the Matsubara frequency of fermions $\omega_m=(2m+1)\pi/\beta$ and the correlation function can be written as:
\begin{equation}
    \Pi_{\mu\nu}(\bq,i\omega_n)=\frac{e^2}{\beta V\hbar^2}\sum_{\bk,m}\text{tr}\left[\cG(\bk+\bq,i\omega_m+i\omega_n)\Xi(\bk+\bq/2)\cG(\bk,i\omega_m)\Xi(\bk+\bq/2)\right].
\end{equation}
One can then take analytical continuation by $i\omega_n\rightarrow\omega$ to arrive at the final form of $\Pi_{\mu\nu}(\bq,\omega)$.

The Green's function can be written under the basis of the energy eigen states denoted by $|a,\bk\rangle$ whose eigen value is $E_{a,\bk}$ as:
\begin{equation}
    \cG(\bk,i\omega_m)=\sum_a\frac{|a,\bk\rangle\langle a,\bk|}{i\omega_m-E_{a,\bk}}.
\end{equation}
The correlation function can then be written as:
\begin{equation}
    \Pi_{\mu\nu}(\bq,i\omega_n=0)=\frac{e^2}{\beta V\hbar^2}\sum_{\bk,m}\sum_{a,b} \frac{1}{i\omega_m-E_{a,\bk+\bq}}\frac{1}{i\omega_m-E_{b,\bk}} \langle a,\bk+\bq|\Xi(\bk+\bq/2)|b,\bk\rangle \langle b,\bk|\Xi(\bk+\bq/2)|a,\bk+\bq\rangle.
\end{equation}
The Matsubara sum can be taken analytically, giving rise to
\begin{equation}
   \Pi_{\mu\nu}(\bq,i\omega_n=0)=\frac{e^2}{V\hbar^2}\sum_{\bk}\sum_{a,b} \frac{n_F(E_{a,\bk+\bq})-n_F(E_{b,\bk})}{E_{a,\bk+\bq}-E_{b,\bk}} \langle a,\bk+\bq|\Xi(\bk+\bq/2)|b,\bk\rangle \langle b,\bk|\Xi(\bk+\bq/2)|a,\bk+\bq\rangle.
\end{equation}

\section{Raman spectra}
If the contribution of surface reflection is neglected, the differential cross-section of Raman spectroscopy can be expressed by the following equation:
\begin{align}
    \frac{\partial^{2}\sigma}{\partial\Omega\partial\omega}\propto-\frac{1}{1-e^{-\beta\omega}}\mathrm{Im}R(\bq, \omega),
\end{align}
in which $R(\bq, \omega)$ is the Raman response function, which in the time domain can be expressed as the correlation function of the Raman density operator $\rho_{\gamma}$:
\begin{align}
    R(\bq, \omega)=-i\theta(t)\langle[\rho_{\gamma}(\bq, t), \rho_{\gamma}(-\bq, 0)]\rangle.
\end{align}
$\rho_{\gamma}$ can be separated into a resonant part and a non-resonant part. Under conditions of low temperature, small momentum, and low incident light frequency, only the contribution from the non-resonant part needs to be considered. After applying the effective mass approximation, $\rho_{\gamma}(\bq, t)$ can be written as:
\begin{align}
    \rho_{\gamma}(\bq, t)=\sum_{\bk,\alpha,\beta,\sigma}\gamma_{\bk}^{\alpha\beta}c_{\bk+\frac{\bq}{2},\alpha,\sigma}^{\dagger}(t)c_{\bk-\frac{\bq}{2},\beta,\sigma}(t),
\end{align}
where $\alpha$, $\beta$ are orbital indices and $\sigma$ is the spin index. $\gamma_{\bk}$ is a matrix and can be written as:
\begin{align}
    \gamma_{\bk}=\sum_{\mu,\nu}e_{\mu}^{I}\frac{\partial^2H}{\partial k_{\mu}\partial k_{\nu}}e_{\nu}^{S},
\end{align}
$e^{I}$ and $e^{S}$ are the polarization vectors of the incident and scattered light, respectively, while $\mu$ and $\nu$ represent the $x$, $y$, $z$ directions in the Cartesian coordinate system. $H$ is the normal state Hamiltonian of the superconductor. Under typical experimental conditions, we only consider the long-wavelength limit $\bq \to 0$ next.

As is common practice, we consider the corresponding Matsubara Green's function in imaginary time:
\begin{align}
    R(0, \tau)=-\langle T_{\tau}(\rho_{\gamma}(0, \tau), \rho_{\gamma}(0,0))\rangle.
\end{align}
Using Wick's theorem, this Raman response function can be written in a compact form:
\begin{align}
    R(0, \tau)=\mathrm{Tr}[\widetilde{\gamma_{\bk}}\cG(\bk, \tau)\widetilde{\gamma_{\bk}}\cG(\bk, -\tau)].  
\end{align}
Here, $\cG(\bk, \tau)$ is the Gor'kov Green's function in the Nambu basis, and $\widetilde{\gamma_{\bk}}$ is the original $\gamma_{\bk}$ matrix extended to the same Nambu basis, which has the following form:
\begin{align}
    \widetilde{\gamma_{\bk}}=\begin{pmatrix}
        \gamma_{\bk} & 0 \\
        0 &-\gamma_{\bk}^{T}
    \end{pmatrix}.
\end{align}
Applying the Fourier transformation to Matsubara’s frequency domain, we have:
\begin{align}
    R(0, i\omega_{n})=\frac{1}{\beta}\sum_{\bk,\omega_{m}}\mathrm{Tr}(\widetilde{\gamma_{\bk}}\cG(\bk, i\omega_{m}+i\omega_{n})\widetilde{\gamma_{\bk}}\cG(\bk, i\omega_{m})),
\end{align}
where $\cG(\bk, i\omega_{m})=(i\omega_{m}-H_{BdG})^{-1}$, $H_{BdG}$ is the Bogoliubov-de Gennes (BdG) Hamiltonian for the superconductor.

It is more convenient to work in eigenvector basis of BdG Hamiltonian when we do the Matsubara’s frequency summation later. Since the BdG Hamiltonian can be diagonalized by a unitary transformation, and assuming this unitary matrix is $U_{\bk}$, we have
\begin{align}
    R(0, i\omega_{n})=\frac{1}{\beta}\sum_{\bk,\omega_{m}}\mathrm{Tr}(U_{\bk}^{\dagger}\widetilde{\gamma_{\bk}}U_{\bk}U_{\bk}^{\dagger}\cG(\bk, i\omega_{m}+i\omega_{n})U_{\bk}U_{\bk}^{\dagger}\widetilde{\gamma_{\bk}}U_{\bk}U_{\bk}^{\dagger}\cG(\bk, i\omega_{m})U_{\bk}).
\end{align}
Let $\Gamma_{\bk}=U_{\bk}^{\dagger}\widetilde{\gamma_{\bk}}U_{\bk}$, expanding the trace operation in the above equation, we obtain:
\begin{align}
    R(0, i\omega_{n})=\frac{1}{\beta}\sum_{\bk,\omega_{m},m,n}\Gamma_{\bk}^{mn}\Gamma_{\bk}^{nm}\frac{1}{i\omega_{n}+i\omega_{m}-E_{n}}\frac{1}{i\omega_{m}-E_{m}},
\end{align}
where $E_{n}$ is the $n^{th}$ eigenvalue of the BdG Hamiltonian. After summing over the Matsubara frequencies and then analytically continuing to the real frequency axis, the final result of the Raman response function can be obtained:
\begin{align}
    R(0, \omega)=\sum_{\bk,\omega_{m},m,n}\Gamma_{\bk}^{mn}\Gamma_{\bk}^{nm}(\frac{1}{e^{\beta E_{m}}+1}-\frac{1}{e^{\beta E_{n}}+1})\frac{1}{\omega+E_{m}-E_{n}+i0^{+}}.
\end{align}
By applying the following formula, the theoretical results of the Raman spectrum can be obtained:
\begin{align}
    \frac{\partial^{2}\sigma}{\partial\Omega\partial\omega}\propto-\frac{1}{1-e^{-\beta \omega}}\mathrm{Im}R(0, \omega).
\end{align}
\end{document}